\UseRawInputEncoding
\documentclass[a4paper,aps,prd,twocolumn,tightenlines,preprintnumbers,nofootinbib,showkeys, superscriptaddress]{revtex4-1}
\usepackage{fullpage}
\usepackage{amsfonts}
\usepackage{amsmath}
\usepackage{slashed}
\usepackage{amssymb}
\usepackage{graphicx}
\usepackage{makeidx}
\usepackage{cancel}
\usepackage{epic}
\usepackage{eepic}
\usepackage{epsfig}
\usepackage{latexsym}
\usepackage[dvipsnames]{xcolor}
\usepackage{float}
\usepackage{multirow}
\usepackage[export]{adjustbox}
\usepackage{xurl,hyperref}
\usepackage{enumitem}
\hypersetup{colorlinks=true,citecolor=red,linkcolor=NavyBlue,urlcolor=NavyBlue}
\usepackage[utf8]{inputenc}
\usepackage[caption=false]{subfig}
 
\usepackage{natbib}
\usepackage{relsize}
\usepackage[left=1.5cm,right=1.5 cm,top=2 cm,bottom=2 cm]{geometry}
\usepackage{mathptmx}
\begin{document}
\relscale{1.05}
\captionsetup[subfigure]{labelformat=empty}

\title{Direct production of SM-singlet scalars at the muon collider}

\author{Bibhabasu De}
\email{bibhabasude@gmail.com}
\affiliation{Department of Physics, The ICFAI University Tripura, Kamalghat-799210, India}

\date{\today}

\begin{abstract}
\noindent
The present work proposes a minimal extension of the Standard Model~(SM) where a gauge-singlet scalar~($\phi$) can be directly produced at the muon colliders without relying on its mixing with any other doublet state present in the theory. The New Physics~(NP) interactions include a TeV-scale scalar leptoquark of electromagnetic charge 1/3 arising naturally in a grand unifying gauge formulation. Within the proposed framework, the SM-singlet scalar can effectively couple to various SM fields at the one-loop level, out of which the $\bar{\mu}\mu\phi$ and $\gamma\gamma\phi$ couplings are crucial to produce it at the future muon colliders. Assuming $\mathcal{O}(1)$ NP couplings, the decay widths and production cross-section of the singlet scalar have been discussed in detail over the considered parameter space. Depending on the resonance scale, di-lepton and/or di-gluon channels can be significant to test/falsify the model. 
\end{abstract}
	
\maketitle	

\section{Introduction}
\noindent
The discovery of the 125 GeV SM-like Higgs boson at the LHC~\cite{ATLAS:2012yve, CMS:2012qbp} led to experimental validation for various SM predictions. Though this observation completed the SM particle spectrum, marked new directions for the Beyond Standard Model~(BSM) physics --- a significant one being the possibility of other scalar fields~\cite{ATLAS:2012tny, CMS:2012fgz, ATLAS:2012ube, ATLAS:2012hi, CMS:2012qms, CMS:2014ccx, ATLAS:2017tlw, CMS:2015hra, ATLAS:2018sbw, CMS:2019bnu, cms2023search}. From the phenomenological perspective, an SM-singlet scalar can be a well-motivated NP candidate to accommodate different BSM observations within a common theoretical framework~\cite{Ghosh:2015apa, De:2021crr, Keus:2017ioh, Lee:2014rba, Bhattacharya:2016qsg, Belanger:2021slj, Sakurai:2022hwh} while presenting the simplest possible explanation for the observed dark matter~(DM) abundance~\cite{McDonald:1993ex, Biswas:2011td, Cline:2013gha, Das:2020ozo, De:2023sqa}. Further, such a gauge-singlet scalar can play a vital role in solving the so-called $\mu$ problem in the Minimal Supersymmetric Standard Model~\cite{Ellwanger:2009dp} and understanding the electroweak~(EW) phase transition~\cite{Espinosa:2011ax, Huber:2000mg, Ham:2004cf}. Several models have been proposed to study the production prospects of an SM-singlet scalar at the LHC, with most of them relying upon its mixing with the Higgs-like doublet states present in the theory. In general, for a non-negligible value of the mixing angle, gluon fusion appears as the leading order production channel. The singlet scalar may also be produced through the cascade decays of the doublet state. However, such formulations, in general, result in a much suppressed production cross-section for the gauge-singlet state. Ref.~\cite{Bhaskar:2020kdr}, for the first time in literature, proposed a direct production mechanism for a real singlet scalar at the LHC without any mixing with the doublet states. The model considered a simple extension of the SM with a scalar leptoquark~(LQ) and a right-handed neutrino where the singlet scalar can be significantly produced through the quark fusion channels with the leading contribution arising at the one-loop level. The idea has been further explored in Ref.~\cite{Bhaskar:2022ygp} in the context of di-Higgs production at the LHC. 

However, to probe the NP in the multi-TeV range, muon colliders~\cite{Delahaye:2019omf, Long:2020wfp, AlAli:2021let, Accettura:2023ked, MuonCollider:2022xlm} can provide with a more precise and clean environment compared to their hadronic counterparts. A crucial difference between hadronic and leptonic colliders is that the hadrons being composite particles, only a fraction of the total beam energy is available for the actual partonic collision while leptons having no substructure the complete center of mass~(COM) energy can be utilized, making it an excellent testing tool for the BSM theories~\cite{Liu:2021jyc, Huang:2021nkl, Han:2020pif, Han:2020uak, Costantini:2020stv, Asadi:2021gah, Liu:2022byu, Bao:2022onq, Ghosh:2023xbj, Jana:2023ogd, Bhattacharya:2023beo}. Further, the synchrotron radiation that effectively constrains the COM energies of circular $e^+e^-$ colliders is suppressed due to the considerable mass of muons, leading to both high energies and high luminosities simultaneously. A muon collider can also be substantial to search for muon-specific NP in the sub-TeV range~\cite{Capdevilla:2021rwo,Buttazzo:2020ibd,Han:2021lnp,Yin:2020afe,Capdevilla:2020qel,Celada:2023oji}. The production prospects of BSM scalars have already been studied in the frame of muon colliders~\cite{Akeroyd:1999xf, Akeroyd:2000zs, Hashemi:2012nz, Hashemi:2012we, Buttazzo:2015bka, Buttazzo:2018qqp, Han:2021udl, AlAli:2021let, Belfkir:2023lot, Ouazghour:2023plc}. In particular, Refs.~\cite{Buttazzo:2015bka, Buttazzo:2018qqp, AlAli:2021let} have discussed the case of a gauge-singlet scalar, but again with a production cross-section that significantly depends on the coupling strength of the singlet with the doublet states considered in the model. 

This paper proposes a simple BSM formulation by extending the SM with a real singlet scalar $\phi$ and a scalar leptoquark $S_1$, transforming as ($\bar{\mathbf{3}}$, {\bf 1}, 1/3) under the SM gauge group. Leptoquarks~(for recent reviews, see Refs.~\cite{Dorsner:2016wpm, Davidson:1993qk, Hewett:1997ce, Nath:2006ut}) are hypothetical bosons that arise naturally in the grand unified theories~(GUT)~\cite{Pati:1974yy, Georgi:1974sy, Georgi:1974my, Fritzsch:1974nn, Kang:2007ib} to couple leptons and quarks at the tree level and can be a remarkable NP candidate to explain various BSM phenomena. For example, several B-meson anomalies can be explained by augmenting the SM with a LQ~\cite{Dorsner:2013tla,Gripaios:2014tna,Becirevic:2015asa,Becirevic:2016yqi, Crivellin:2017zlb,Cline:2017aed,DiLuzio:2017chi,Mandal:2018kau,Aydemir:2019ynb,Crivellin:2019dwb,DaRold:2019fiw}. LQs may also be vital for studying the DM phenomenology~\cite{Mandal:2018czf,Choi:2018stw,Mohamadnejad:2019wqb} and the leptonic observables~\cite{Djouadi:1989md,Couture:1995he,Cheung:2001ip,Dorsner:2019itg,Greljo:2021xmg,Kowalska:2018ulj,Athron:2021iuf, Bhaskar:2022vgk,Parashar:2022wrd, Mandal:2019gff,Ghosh:2022vpb, De:2023acg}. Refs.~\cite{Bhaskar:2020kdr,Bhaskar:2022ygp,DaRold:2021pgn,Agrawal:1999bk,Enkhbat:2013oba} have discussed the significance of LQ in the context of producing scalar particles.
Note that, though the simplest GUT models need to assume a heavy LQ~\cite{Super-Kamiokande:2014otb, Dorsner:2012nq} which is beyond the reach of current and future colliders, there exist GUT formulations that can explain the stability of proton with a TeV-scale scalar LQ~\cite{BUCHMULLER1986377,Murayama:1991ah,Dorsner:2005fq, GEORGI1979297, FileviezPerez:2007bcw, Senjanovic:1982ex}. In this paper, the latter GUT formalism will be considered as the governing gauge theory to describe the NP interactions.  

In contrast to the currently available muon collider studies, within the proposed framework, the SM-singlet scalar $\phi$ can couple to the muons and photons at one-loop level, leading to a direct production of $\phi$ at the muon colliders without mixing with any additional doublet state. Though Ref.~\cite{Bhaskar:2020kdr} has considered a similar formulation, it demands a right-handed neutrino in addition to $S_1$ for having a significant production cross-section of $\phi$ at the LHC. Thus, with the requirement of an additional right-handed neutrino being removed, the present model can be treated as the most minimal phenomenological construction to produce $\phi$ directly at the colliders. Note that the core target of this work is to propose a direct production mechanism for $\phi$ and not to present a detailed collider analysis. Thus, to study the production prospects of $\phi$, only the leading order contributions have been considered and plotted in general. However, to illustrate the effect of higher order corrections, a few benchmark values have been listed considering the Beam Energy Spread~(BES) and Initial State Radiation~(ISR). They effectively modify the sharp Breit-Wigner nature of the hard collision cross-section~\cite{Greco:2016izi, Franceschini:2021aqd} and could be crucial for a narrow-width resonance.
   
The rest of the paper has been organized as follows. Sec.~\ref{sec:1} defines the NP interactions when the SM is extended to the scale of $\mathcal{O}(1)$ TeV. Sec.~\ref{sec:2} presents the effective couplings and decay widths of the real gauge-singlet scalar $\phi$ to various SM fields in the presence of a scalar LQ $S_1$, followed by the calculation of the total production cross-section of $\phi$ at the muon collider. The detection prospects of $\phi$ have been discussed in Sec.~\ref{sec:3}, and finally, the work has been summarized in Sec.~\ref{sec:4}. 

\section{The Model}
\label{sec:1}
\noindent
The model proposes a {\it minimal} extension of the SM to a NP scale $\Lambda_{\rm NP}\sim \mathcal{O}(1)$ TeV, where the observed particle spectrum can be augmented with a scalar LQ $S_1$ and a real SM-singlet scalar $\phi$. As discussed in the Introduction, considering such a TeV-scale scalar LQ is well-motivated from the perspective of various GUT models. Table~\ref{tab:parti} enlists the complete particle spectrum of the assumed framework along with their transformations under the SM gauge group $\mathcal{G}_{\rm SM}\supset SU(3)_C\times SU(2)_L\times U(1)_Y$. 
\begin{table}[!ht]
\begin{tabular}{|c|c|c|}
\hline
Fields & Generations & $SU(3)_C\times SU(2)_L\times U(1)_Y$ \\
\hline
\hline
$L_L=(\nu_L\quad \ell_L)^T$  & 3 & ({\bf 1}, {\bf 2}, -1/2)  \\
		$\ell_R= (e_R,\,\mu_R,\,\tau_R)$ & 3 & ({\bf 1}, {\bf 1}, -1) \\
		$Q_L=(u_L\quad d_L)^T$  & 3 & ({\bf 3}, {\bf 2}, 1/6) \\
		$U_R=(u_R,\,c_R,\,t_R)$  & 3 & ({\bf 3}, {\bf 1}, 2/3) \\
		$D_R=(d_R,\,s_R,\,b_R)$  & 3 & ({\bf 3}, {\bf 1}, -1/3) \\
		$H = (H^+ \quad H^0)^T $ & 1 & ({\bf 1}, {\bf 2}, 1/2)  \\
\hline
		$S_1$ & 1 & ($\bar{\mathbf{3}}$, {\bf 1}, 1/3)	\\
		$\phi$ & 1 & ({\bf 1}, {\bf 1}, 0)	\\
		\hline
\end{tabular} 
\caption{Complete particle spectrum at the NP scale $\Lambda_{\rm NP}$ with their respective gauge transformations under $\mathcal{G}_{\rm SM}$. The electromagnetic~(EM) charge can be defined as $Q_{\rm EM}=T_3+Y$.}
\label{tab:parti}
\end{table}

The model can be described through the Lagrangian,
\begin{align*}
\mathcal{L}=\mathcal{L}_{\rm SM}+\mathcal{L}_{\rm NP},
\end{align*}
where, $\mathcal{L}_{\rm SM}$ encapsulates the SM interactions only, and
\begin{align}
\mathcal{L}_{\rm NP}=~& (D^\mu S_1)^\dagger(D_\mu S_1)+\frac{1}{2}\partial^2\phi \nonumber\\
&-\Big[\xi_L^{ij} (\bar{Q}_L^{Cia}\epsilon^{ab}L_L^{jb})S_1+ \xi_R^{ij} 
(\bar{U}_R^{Ci}\ell_R^j)S_1+{\rm h.c.}\Big]\nonumber\\
&\qquad\qquad\qquad\qquad\qquad\qquad-\mathcal{V}(H,\,S_1,\,\phi)\nonumber\\
=~& (D^\mu S_1)^\dagger(D_\mu S_1)+\frac{1}{2}\partial^2\phi-\Big[\left(\bar{u}_L^{Ci}\xi_L^{ij} \ell_L^{j}\right)S_1\nonumber\\
&-\left(\bar{d}_L^{Ci}\xi_L^{ij} \nu_L^{j}\right)S_1+ \xi_R^{ij} 
(\bar{U}_R^{Ci}\ell_R^j)S_1+{\rm h.c.}\Bigg]\nonumber\\
&\qquad\qquad\qquad\qquad\qquad\qquad-\mathcal{V}(H,\,S_1,\,\phi).
\label{eq:NP_Lag}
\end{align}
Here, the color indices have been suppressed. The superscript $C$ marks the charge conjugate state; $\{a,b\}$ and $\{i,j\}$ represent the $SU (2)_L$ and flavor indices, respectively. For simplicity, mixing among the quarks has been ignored in Eq.~\eqref{eq:NP_Lag}, i.e., $\mathbb{V}_{\rm CKM}=\mathbb{I}_{3\times 3}$. $D^\mu$ denotes the covariant derivative corresponding to the LQ $S_1$. The scalar potential can be defined as,
	\begin{align}
\mathcal{V}(H,\,& S_1,\,\phi)= \lambda \left(H^\dagger H\right) \left(S^\dag_1S_1\right)+\beta\phi 
\left(S^\dag_1S_1\right)+\lambda_1 (H^\dagger H)\phi^2\nonumber\\
& +\rho\phi(H^\dagger H)+\lambda_2\phi^2 
\left(S^\dag_1S_1\right)+ \frac{1}{2}M^2_\phi \phi^2 + M^{2}_{S_1}\left(S^\dag_1S_1\right).
	\label{eq:pot}
	\end{align}
However, to ensure that the presence $\phi$ doesn't affect the production or decay of the SM Higgs, one can assume $\lambda_1$ and $\rho$ to be negligible. Further, to evade any additional constraint on the parameter space arising due to the correction of the trilinear Higgs coupling, it's beneficial to set $\lambda\to 0$. Note that $\beta$ is a mass dimensional coupling which should, in principle, represent the highest scale of the theory. Moreover, it plays a crucial role in producing a single $\phi$ at the muon collider. After electroweak symmetry breaking~(EWSB) only the SM Higgs acquires a vacuum expectation value~(VEV) $v=246$ GeV, such that,
	 \begin{align}
H = 
\frac{1}{\sqrt{2}}\begin{pmatrix}
0 \\
v + h
\end{pmatrix}. 
\end{align}
The physical masses of all the SM fermions can be defined through their respective Yukawa interactions. Note that, in the limit $\lambda\to 0$, $M_{S_1}$ represents the physical mass of $S_1$.
\section{Decay and Production of $\phi$}
\label{sec:2}
\noindent
In the presence of $S_1$, the singlet scalar $\phi$ can couple to the SM leptons, up-type quarks, gluons, and photons at one-loop level. However, at the muon collider, the production cross-section of $\phi$ majorly depends on the effective $\bar{\mu}\mu\phi$ coupling generated through Fig.~\ref{fig:mu_phi} as shown below.
   \begin{figure}[!ht]
   \centering
   \includegraphics[scale=0.5]{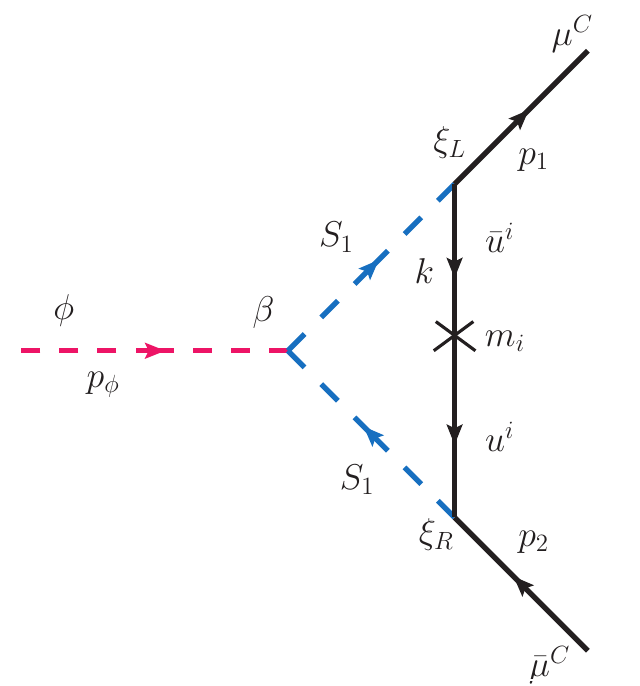}
   \caption{Leading order contribution to $\phi\to \bar{\mu}\mu$. Here, $k$ denotes the loop momentum, and $p_\phi=p_1-p_2$ defines momentum conservation.}
   \label{fig:mu_phi}
   \end{figure}
   
The effective coupling corresponding to Fig.~\ref{fig:mu_phi} can be cast as,
\begin{align}
-iY_{\bar{\mu}\mu\phi}&= \sum_{i=\,u,c,t}(\xi_L^{\mu i})^*\xi_R^{i\mu}\beta m_i\int\frac{d^4k}{(2\pi)^4} \Bigg[\frac{(k^2+m_i^2)}{(k^2-m_i^2)^2}\times\nonumber\\
&\qquad\qquad\frac{1}{\{(k+p_1)^2-M_{S_1}^2\}\{(k+p_2)^2-M_{S_1}^2\}}\Bigg]\nonumber\\
=& \sum_{i=\,u,c,t}(\xi_L^{\mu i})^*\xi_R^{i\mu}\beta m_i\int\frac{d^4k}{(2\pi)^4} \Bigg[\nonumber\\
&\frac{1}{(k^2-m_i^2)\{(k+p_1)^2-M_{S_1}^2\}\{(k+p_2)^2-M_{S_1}^2\}}\nonumber\\
&+\frac{2m_i^2}{(k^2-m_i^2)^2\{(k+p_1)^2-M_{S_1}^2\}\{(k+p_2)^2-M_{S_1}^2\}}\Bigg]\nonumber\\
=& \sum_{i=\,u,c,t}\left(\frac{i\,(\xi_L^{\mu i})^*\xi_R^{i\mu}\beta m_i}{16\pi^2}\right)\Bigg[\mathcal{I}_1+2m_i^2\times \mathcal{I}_2\Bigg],
\label{eq:Y1}
\end{align}
where
\begin{align}
\mathcal{I}_1=&\ -C_0\left(0,M_\phi^2,0, m_i^2, M_{S_1}^2,M_{S_1}^2\right),\nonumber\\
\mathcal{I}_2=&\ D_0\left(0,0,M_\phi^2,0,0,0, m_i^2,m_i^2, M_{S_1}^2,M_{S_1}^2\right).
\end{align}

To obtain the Eq.~\eqref{eq:Y1}, an on-shell decay of $\phi$ has been considered so that $p_\phi^2=(p_1-p_2)^2=M_\phi^2$. Further, one can easily neglect the muon mass with respect to the BSM scale involved in the theory, leading to the physically acceptable approximation $M_\phi^2\simeq -2p_1.p_2$. $C_0$ and $D_0$ define the standard Passarino-Veltmen functions~\cite{Romao:2020} for the scalar 3-point and 4-point one-loop integrals, respectively. From here on, the flavor index will be suppressed for the LQ couplings, and it will be considered as $|\xi|^2=(\xi_L^{\mu i})^*\xi_R^{i\mu}$. Note that one can always use the flavor-specific values of $\xi^{i\mu}_{L,R}$, constrained through various low-energy lepton phenomena~\cite{Mandal:2019gff}, to enhance the {\it phenomenological accuracy} of the results, but that will not alter the physics significantly apart from an overall scaling of the observables presented here. Thus, this paper will consider $|\xi|^2=1$ and $\beta=4$ TeV for all the numerical illustrations.
\begin{figure}[!ht]
\centering
\includegraphics[scale=0.6]{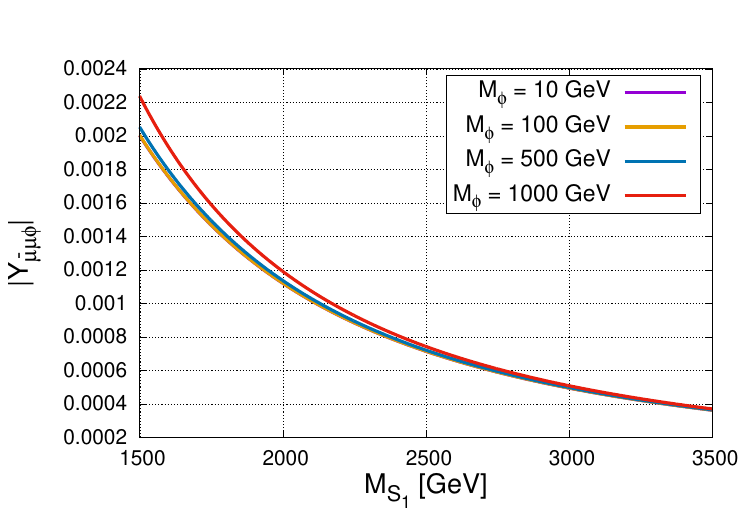}
\caption{Variation of $|Y_{\bar{\mu}\mu\phi}|$ as a function of $M_{S_1}$ for $M_\phi=10$ GeV~(violet), 100 GeV~(golden), 500 GeV~(sky), and 1000 GeV~(red).}
\label{fig:Y_phi_mu}
\end{figure}
Fig.~\ref{fig:Y_phi_mu} shows the variation of the effective $\bar{\mu}\mu\phi$ coupling as a function of the LQ mass $M_{S_1}$. Further, to study the effect of $M_\phi$ on $Y_{\bar{\mu}\mu\phi}$, four illustrative values of the scalar mass have been chosen at 10 GeV, 100 GeV, 500 GeV, and 1000 GeV. However, for the smaller values of $(M_\phi/M_{S_1})^2$, the mass scale of $\phi$ becomes irrelevent. 
\subsection{Decay Width}
\label{sec:dec}
\noindent
Since the proposed model allows $\phi$ to couple with various SM fields, the corresponding decay widths of $\phi$ can be calculated in terms of its effective couplings within a kinematically permitted region of the parameter space. Thus, assuming a COM frame of reference, the decay widths for $\phi\to \bar{\ell}\ell$ and $\phi\to \bar{u}u$ can be defined as,
\begin{align}
\Gamma_{\phi\to \bar{\ell}\ell}&\ =\frac{\mathcal{N}_C}{8\pi M_\phi^2}|Y_{\bar{\ell}\ell\phi}|^2\left(M_\phi^2-4m_\ell^2\right)^{3/2},\nonumber\\
\Gamma_{\phi\to \bar{u}u}&\ =\frac{\mathcal{N}_C}{8\pi M_\phi^2}|Y_{\bar{u}u\phi}|^2\left(M_\phi^2-4m_u^2\right)^{3/2},
\end{align}
where, $\mathcal{N}_C=3$ labels the color degeneracy factor and $u$ symbolizes the up-type quarks. Note that, $Y_{\bar{\ell}\ell\phi}$ is same as $Y_{\bar{\mu}\mu\phi}$ since to obtain Eq.~\eqref{eq:Y1}, the external lepton masses have been ignored. Further, the effective coupling between $\phi$ and the up-type quarks can be formulated as,
\begin{align}
-iY_{\bar{u}u\phi}=& \sum_{\ell=\,e,\mu,\tau}\left(\frac{i\,|\xi|^2\beta m_\ell}{16\pi^2}\right)\Bigg[-C_0\left(0,M_\phi^2,0, m_\ell^2, M_{S_1}^2,M_{S_1}^2\right)\nonumber\\
&+2m_\ell^2\times D_0\left(0,0,M_\phi^2,0,0,0, m_\ell^2,m_\ell^2, M_{S_1}^2,M_{S_1}^2\right)\Bigg],
\label{eq:Y2}
\end{align}
where the external quark masses have been dropped in comparison to $\Lambda_{\rm NP}$. Note that the calculational steps to reach Eq.~\eqref{eq:Y2} are the same as those followed for Eq.~\eqref{eq:Y1}.

\begin{figure}[!ht]
\includegraphics[scale=0.5]{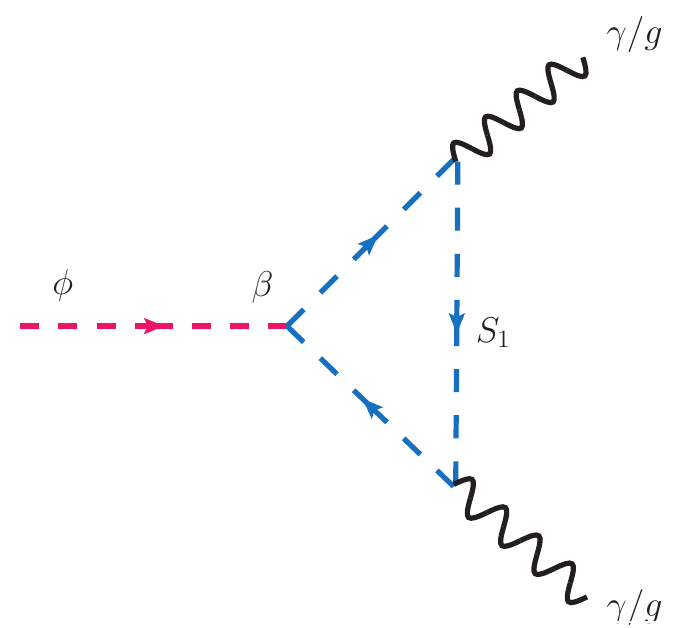}
\vspace{0.3 cm}\\
\includegraphics[scale=0.5]{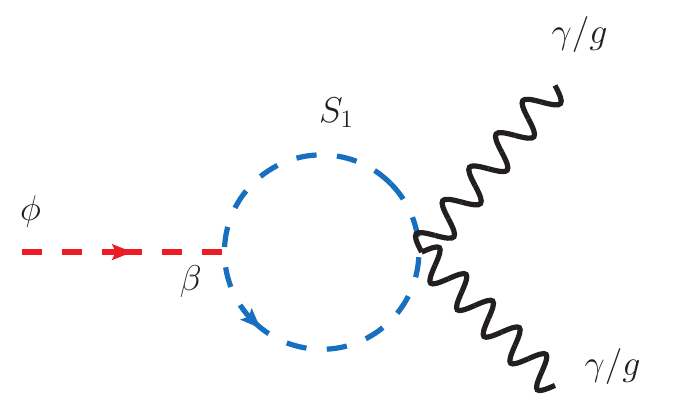}
\caption{Leading order contributions to $\phi\to \gamma\gamma\,(gg)$ processes.}
\label{fig:phi_gg}
\end{figure}

In the presence of a color-triplet charged scalar $S_1$, $\phi$ can also decay through the di-gluon and di-photon channels as depicted in Fig.~\ref{fig:phi_gg}.

The corresponding decay widths are given by~\cite{Bhaskar:2020kdr,Dorsner:2016wpm},
\begin{align}
\Gamma_{\phi\rightarrow gg}&\ =\ \frac{\mathcal{G}_{\rm F}\alpha_S^2M_\phi^3}{64\sqrt{2}\pi^3}\left|\frac{\beta v }{2M_{S_1}^2}\mathcal{F}\left(\frac{M_\phi^2}{4M_{S_1}^2}\right)\right|^2, \nonumber\\
\Gamma_{\phi\rightarrow\gamma\gamma}&\ =\ \frac{\mathcal{G}_{\rm F}\alpha_{\rm EM}^2M_\phi^3}{128\sqrt{2}\pi^3}\left|\frac{\beta v }{6M_{S_1}^2}\mathcal{F}\left(\frac{M_\phi^2}{4M_{S_1}^2}\right)\right|^2,
\end{align}
where, $\mathcal{G}_{\rm F}$ is the Fermi constant, $\alpha_S$ and $\alpha_{\rm EM}$ define the strong and electromagnetic coupling constants, respectively. The function $\mathcal{F}$ can be defined as,
\begin{align}
  \mathcal{F}(x)=-\frac{[x-\theta(x)]}{x^2}\,,
 \end{align}
 where,
 \begin{align}
 \theta(x)&=\Bigg\{\begin{array}{cc}
 {\rm Arcsin}^2(\sqrt{x}), & x\leq 1\\
 -\frac{1}{4}\left[{\rm ln}\left(\frac{1+\sqrt{1-x^{-1}}}{1-\sqrt{1-x^{-1}}}\right)-i\pi\right]^2, & x>1
 \end{array}
 \label{eq:glu_func}
\end{align}
Therefore, the total decay width of $\phi$ can be cast as, 
\begin{align}
\Gamma_\phi=&\ \sum_{\ell=e,\mu,\tau}\Gamma_{\phi\to \bar{\ell}\ell}+\sum_{u=u,c,t}\Gamma_{\phi\to \bar{u}u}+\Gamma_{\phi\rightarrow gg}+\Gamma_{\phi\rightarrow\gamma\gamma}.
\end{align}
Any other possible decay mode will be extremely suppressed and, thus, can safely be neglected.
\begin{figure}[!ht]
\centering
\subfloat[\qquad(a)]{\includegraphics[scale=0.55]{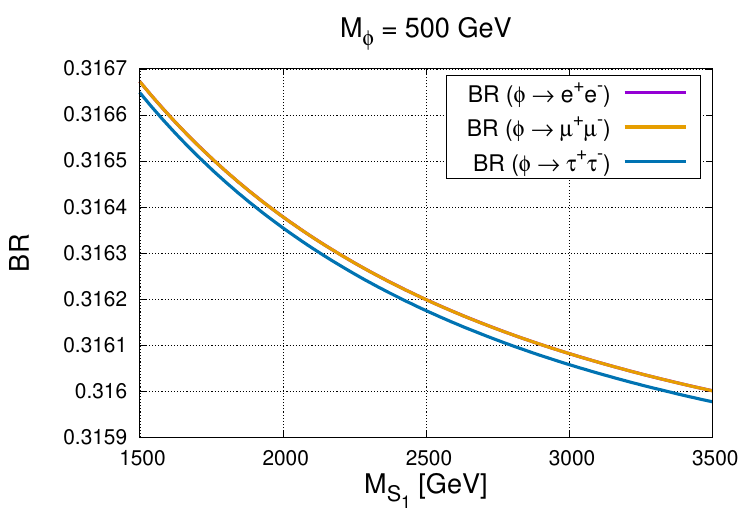}}\\
\subfloat[\qquad(b)]{\includegraphics[scale=0.55]{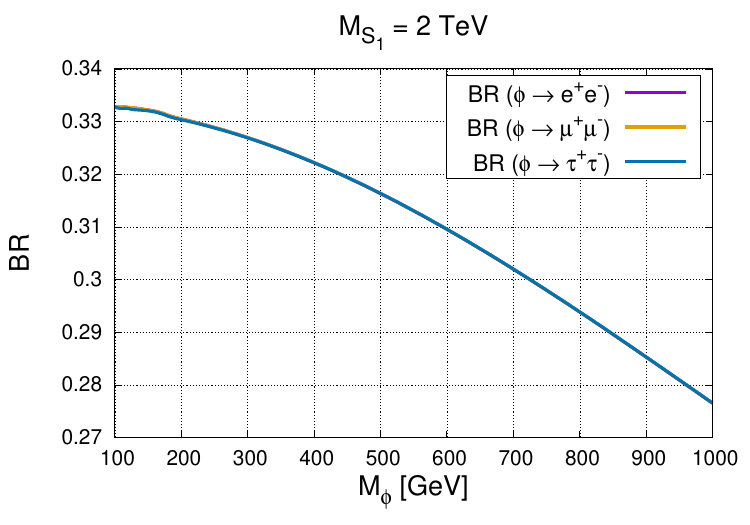}}\\
\subfloat[\qquad(c)]{\includegraphics[scale=0.55]{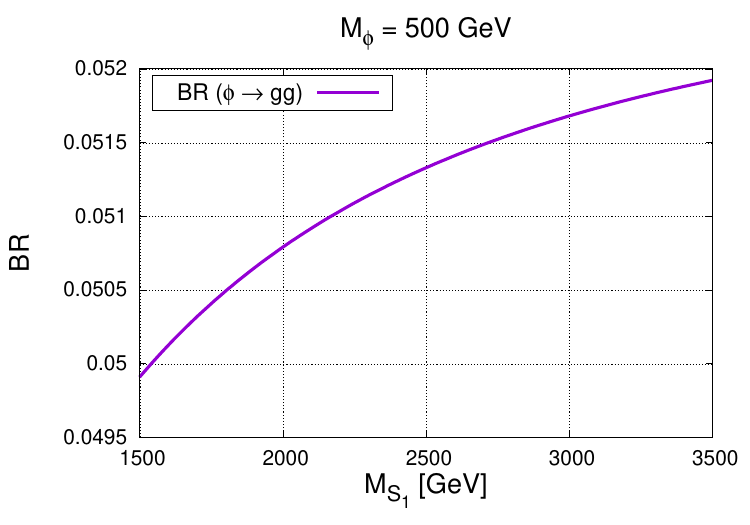}}\\
\subfloat[\qquad(d)]{\includegraphics[scale=0.55]{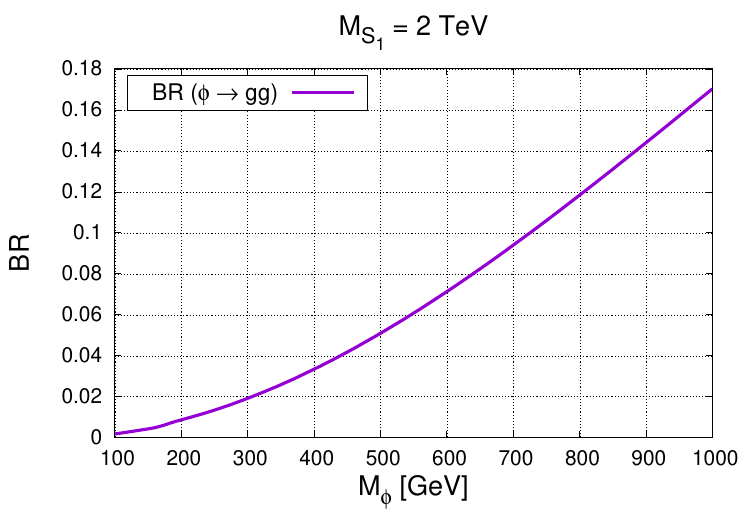}}
\caption{Variation of BR$(\phi\to \ell\ell)$ and BR$(\phi\to gg)$ as a function of $M_{S_1}$ with $M_\phi=500$ GeV~[(a), (c)], and $M_\phi$ for $M_{S_1}$ fixed at 2 TeV~[(b), (d)].}
\label{fig:BR}
\end{figure}
Defining the branching ratio~(BR) for $\phi\to XX$~($X\equiv g,\gamma,e,\mu,\tau,u,c,t$) as BR$(\phi\to XX)=\Gamma_{\phi\to XX}/\Gamma_\phi$, one can numerically determine the most significant decay channel(s) for $\phi$. Table~\ref{tab:BR} enlists the BRs of $\phi$ for a few benchmark values of $M_\phi$ with $M_{S_1}$ being fixed at 2 TeV.
\begin{table}[!ht]
\centering
\begin{tabular}{|c|c|c|c|}
\hline
BR & $M_\phi=100$ GeV & $M_\phi=500$ GeV & $M_\phi=1000$ GeV\\
\hline\hline
BR$(\phi\to \ell^+\ell^-)$ & $3.3\times 10^{-1}$ & $3.2\times 10^{-1}$ & $2.8\times 10^{-1}$\\
$[\ell=e,\mu,\tau]$ & & & \\
\hline
BR$(\phi\to u^+u^-)$, & $3.8\times 10^{-5}$ & $3.6\times 10^{-5}$ & $3.2\times 10^{-5}$\\
$[u=u,\, c]$ & & & \\
\hline
BR$(\phi\to t^+t^-)$ & --- & $1.4\times 10^{-5}$ & $2.6\times 10^{-5}$\\
\hline
BR$(\phi\to g g)$ & $2.1\times 10^{-3}$ & $5.1\times 10^{-2}$ & $1.7\times 10^{-1}$ \\
\hline
BR$(\phi\to \gamma \gamma)$ & $4.5\times 10^{-7}$ & $1.1\times 10^{-5}$ & $3.6\times 10^{-5}$ \\
\hline
\end{tabular}
\caption{BRs for various decay channels of $\phi$ with $M_{S_1}=2$ TeV.}
\label{tab:BR}
\end{table}

Moreover, di-lepton and di-gluon channels being the most relevant ones for the decay of $\phi$, Fig.~\ref{fig:BR}, for completeness, shows a graphical variation of the corresponding BRs.  
Fig.~\ref{fig:BR}\,(a) and \ref{fig:BR}\,(b) depict the variation of BRs of $\phi$ for di-leptonic decay modes as a function of $M_{S_1}$ and $M_\phi$, respectively. In the former case $M_\phi$ has been fixed at 500 GeV while the latter assumes $M_{S_1}=2$ TeV. Fig.~\ref{fig:BR}\,(c) and \ref{fig:BR}\,(d) repeat the same analysis for the di-gluon channel. Note that, for a fixed value of $M_{S_1}$, BR($\phi\to gg$) increases with increasing $M_\phi$, whereas BR($\phi\to \ell^+\ell^-$) shows a decreasing nature. Thus, for the considered numerical set-up one obtains BR($\phi\to gg$) $=$ BR($\phi\to \ell^+\ell^-$) at $M_\phi\approx 1250$ GeV with $M_{S_1}=2$ TeV.
\subsection{Production Cross-Section}
\begin{figure}[!ht]
\centering
\subfloat[\quad(a)]{\includegraphics[scale=0.55]{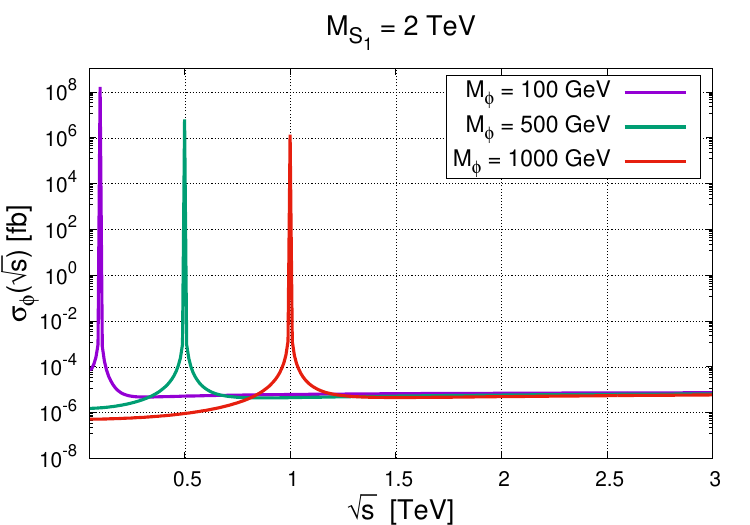}}\\
\subfloat[\quad(b)]{\includegraphics[scale=0.55]{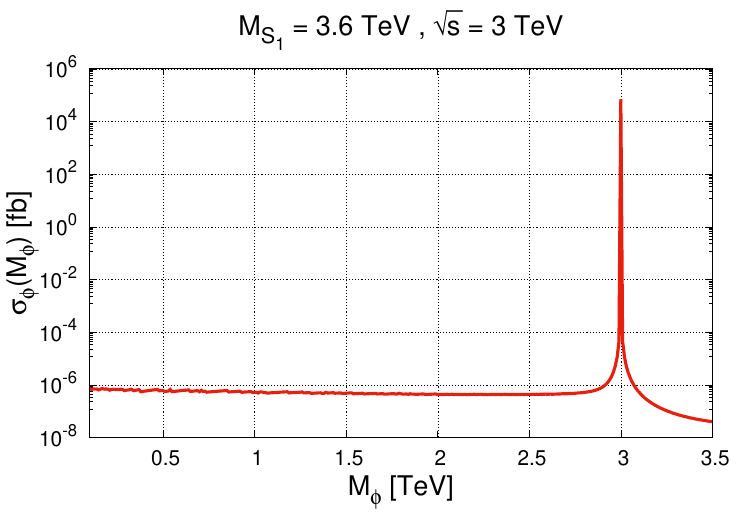}}
\caption{Variation of the production cross-section of $\phi$ as a function of the (a) COM energy for $M_\phi=100$ GeV~(violet), 500 GeV~(green), and 1000 GeV~(red), and (b) $M_\phi$ for $\sqrt{s}=3$ TeV.}
\label{fig:CS}
\end{figure}

\noindent
In a muon collider, the singlet scalar $\phi$ can directly be produced through the two channels: $(i)$ muon fusion and $(ii)$ the photon fusion. In the former case, $\phi$ is directly produced through the fusion of two muons, while for the latter, photons radiated from the colliding muon beams may combine to produce $\phi$ through the effective interaction shown in Fig.~\ref{fig:phi_gg}. However, there is a significant difference between the COM energies. For the muon fusion, it is very close to the collider energy~($\sqrt{s}$) at the leading order, whereas in the case of photon fusion, the COM energy~($\sqrt{s_\gamma}$) is characterized by the effective photon distribution function~(PDF) $\mathcal{P}_\gamma(s_\gamma)$. Thus, the leading order production cross-section of $\phi$ can be formulated as, 
\begin{align}
\sigma_\phi=&~\sigma(\bar{\mu}\mu\to \phi)+\sigma(\gamma\gamma\to \phi)\nonumber\\
=&~4\pi\Bigg[\frac{{\rm BR}(\phi\to \bar{\mu}\mu)}{\mathcal{E}_\mu^2+M_\phi^2}\Bigg]\nonumber\\
&+\left(\frac{4\pi}{s}\right)\int\,\mathcal{P}_\gamma(s_\gamma)\Bigg[\frac{{\rm BR}(\phi\to \gamma\gamma)}{\mathcal{E}_\gamma^2+M_\phi^2}\Bigg]\, ds_\gamma\,,
\label{eq:prod_CS}
\end{align}
where, $\mathcal{E}_\mu=(s-M^2_\phi)/\Gamma_\phi$, and $\mathcal{E}_\gamma=(s_\gamma-M^2_\phi)/\Gamma_\phi$. For the radiated photons,
\begin{align}
\mathcal{P}_\gamma(s_\gamma)=\int_{s_\gamma/s}^1\Bigg[f_\gamma(x)f_\gamma\left(\frac{s_\gamma}{x\,s}\right)\Bigg]\,\frac{dx}{x}\,.
\end{align}
The photon distribution in the muon is parametrized by~\cite{vonWeizsacker:1934nji, PhysRev.45.729, Frixione:1993yw},
\begin{align}
f_\gamma(x)= \frac{\alpha_{\rm EM}}{2\pi}\left(\frac{2+x^2-2x}{x}\right)\,\ln\left(\frac{s}{m_\mu^2}\right)\,,
\end{align}
Fig.~\ref{fig:CS}\,(a) shows the variation $\sigma_\phi$ as a function of the COM energy of the colliding muon beams for three different values of $M_\phi$, with the $M_{S_1}$ being fixed at 2 TeV.
Fig.~\ref{fig:CS}\,(b) displays the variation of $\sigma_\phi$ as a function of the scalar mass $M_\phi$ in a muon collider running at 3 TeV. Though the muon fusion contributes dominantly at resonance, for $\sqrt{s}>M_\phi$, photon fusion becomes more significant despite the coupling and PDF suppressions. To illustrate the significance of photon fusion, one can define a parameter $R_{\gamma/\mu}$ as,
\begin{align}
R_{\gamma/\mu}=\frac{\sigma(\gamma\gamma\to \phi)}{\sigma(\bar{\mu}\mu\to \phi)}~.
\end{align}
\begin{figure}[!ht]
\centering
\includegraphics[scale=0.55]{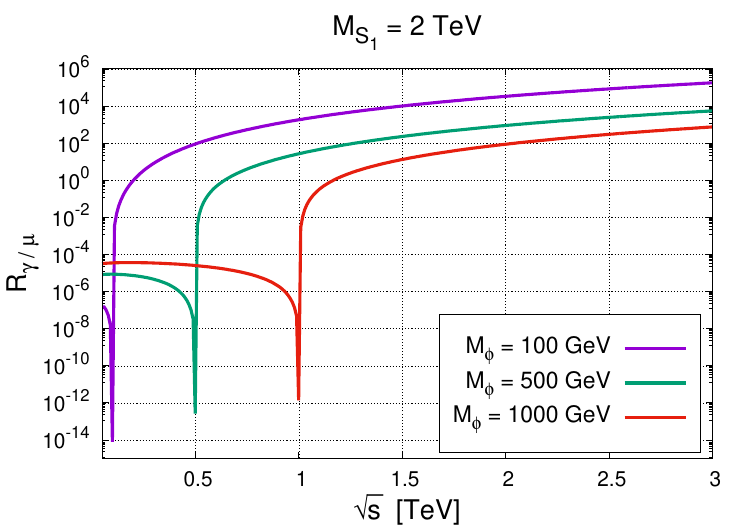}
\caption{Variation of $R_{\gamma/\mu}$ with the beam energy $\sqrt{s}$ for $M_\phi=100$ GeV~(violet), 500 GeV~(green), and 1000 GeV~(red), with $M_{S_1}=2$ TeV.}
\label{fig:Rat}
\end{figure}
It is evident from Fig.~\ref{fig:Rat} that photon fusion can be vital to produce $\phi$ at the muon colliders when the initial beam energy is higher than the scalar mass. Moreover, for a fixed $\sqrt{s}$, the ratio $R_{\gamma/\mu}$ increases with decreasing $M_\phi$, indicating the importance of photon fusion for lighter SM-singlet scalars.
\section{Detection Prospects}
\label{sec:3}
\noindent
As mentioned in Sec.~\ref{sec:dec}, the singlet scalar $\phi$ produced at the muon collider can decay to the SM leptons, up-type quarks, gluons, and photons in the presence of $S_1$ at one-loop level. However, Table~\ref{tab:BR} clearly suggests that a notable signal can only be obtained through the di-lepton and di-gluon channels, with the di-quark and di-photon modes being highly suppressed. In general, the detection cross-section of $\phi$ at the leading order can be cast as,
\begin{align}
\sigma_{XX}=\sigma_\phi\times {\rm BR}(\phi\to XX),
\label{eq:detect}
\end{align}
where, $\sigma_\phi$ is defined through Eq.~\eqref{eq:prod_CS}, and the subscript $XX$ denotes all the possible decay modes of $\phi$, i.e., $X=\ell,\, u,\, \gamma,\, g$.
\begin{figure}[!ht]
\centering
\subfloat[\quad(a)]{\includegraphics[scale=0.55]{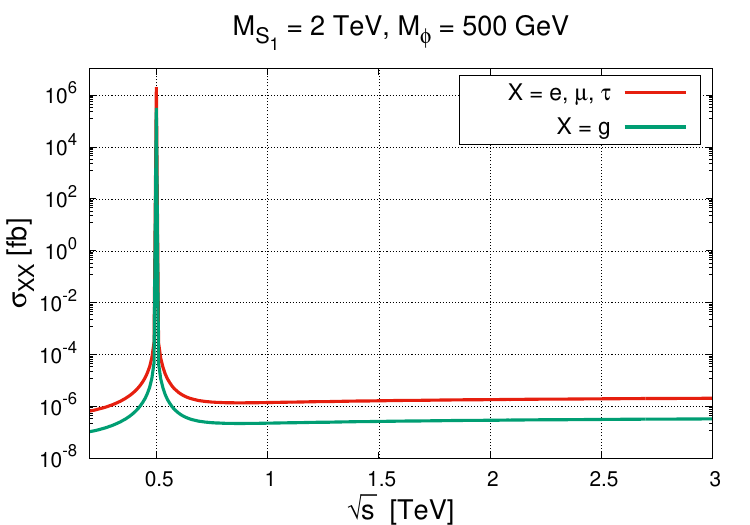}}\\
\subfloat[\quad(b)]{\includegraphics[scale=0.55]{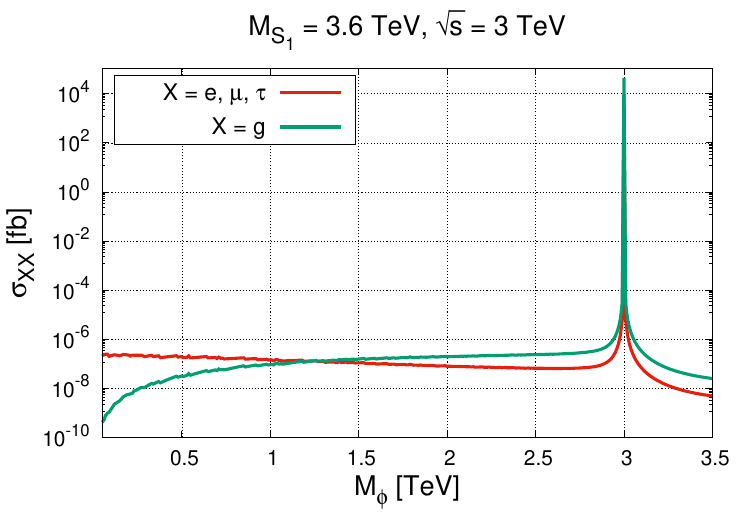}}
\caption{Variation of $\sigma_{XX}$ as a function of (a) the COM energy for $M_\phi=500$ GeV when $M_{S_1}=2$ TeV, and (b) $M_\phi$ with $\sqrt{s}=3$ TeV, and $M_{S_1}=3.6$ TeV.}
\label{fig:CS_BR}
\end{figure}
Fig.~\ref{fig:CS_BR}\,(a) depicts the variation of $\sigma_{XX}$ for the di-lepton~(red) and di-gluon~(green) modes as a function of $\sqrt{s}$ with $M_\phi=500$ GeV, whereas the variation of $\sigma_{XX}$ as a function of the scalar mass $M_\phi$ has been shown in Fig.~\ref{fig:CS_BR}\,(b) for the COM energy being fixed at 3 TeV. Though for $M_\phi\sim\mathcal{O}(100)$ GeV the di-lepton channels show a better detection prospect compared to the di-gluon mode, the order of dominance changes as $M_\phi$ exceeds 1250 GeV. Thus, for a heavy singlet scalar, the di-gluon channel can be more promising. 
\subsection{Effect of ISR and BES}
As stated in the Introduction, Eq.~\eqref{eq:detect} must be convoluted with the ISR probability distribution function and the BES to estimate the actual observable cross-section. Thus, the effective detection cross-section of $\phi$ can be defined as,
\begin{align}
\sigma^{\rm Eff}_{XX}(\sqrt{s})=\int\, d\sqrt{\hat{s}}\,\left(\frac{d\mathbb{L}(\sqrt{s})}{d\sqrt{\hat{s}}}\right)\,\int \mathcal{P}_{\mu\mu}^{\rm ISR}(y,\hat{s})\, \sigma_{XX}\left(y\sqrt{\hat{s}}\right)\, dy\, .
\end{align}
Here, $y$ is the ratio of energy available for hard collision to the COM energy of the muon collider, and $\mathcal{P}_{\mu\mu}^{\rm ISR}$ represents the corresponding ISR profile. $\mathbb{L}$ defines the flux of COM energy with a distribution~\cite{Franceschini:2021aqd},
\begin{align}
\frac{d\mathbb{L}(\sqrt{s})}{d\sqrt{\hat{s}}}=\frac{1}{\sqrt{2\pi\Delta}}.\exp\Bigg[\frac{-\left(\sqrt{\hat{s}}-\sqrt{s}\right)^2}{2\Delta^2}\Bigg],
\label{eq:flux}
\end{align}
where $\Delta=\mathcal{R}\,\sqrt{s/2}$ is the Gaussian energy spread, and $\mathcal{R}$ denotes the percentage beam energy resolution. Though the detailed computation is beyond the scope of this paper, Table~\ref{tab:det} enlists a few illustrative values of the effective detection cross-section $\sigma^{\rm Eff}_{XX}(\sqrt{s})$, obtained with the lowest order approximation of $\mathcal{P}_{\mu\mu}^{\rm ISR}$~\cite{Greco:1975rm}. However, one can find a list of various analytical forms for $\mathcal{P}_{\mu\mu}^{\rm ISR}$ in Ref.~\cite{Greco:2016izi}, which might be useful for a more accurate study.
\begin{table}[!ht]
\centering
\begin{tabular}{|c|c|c|}
\hline
$\left(M_{S_1},\, M_\phi\right)$ & $\sigma^{\rm Eff}_{\ell\ell}\left(\sqrt{s}=M_\phi\right)$ & $\sigma^{\rm Eff}_{gg}\left(\sqrt{s}=M_\phi\right)$\\
\hline\hline
(2.0 TeV, 0.5 TeV) & $2.42\times 10^5$ fb & $3.89\times 10^4$ fb\\
\hline
(2.0 TeV, 1.0 TeV) & $3.04\times 10^4$ fb & $1.87\times 10^4$ fb\\
\hline
(3.6 TeV, 3.0 TeV) & $3.32\times 10^2$ fb & $1.54\times 10^3$ fb\\
\hline
\end{tabular}
\caption{Effective detection cross-section of $\phi$ through the di-lepton and di-gluon channels at resonance for different $M_{S_1}$ and $M_\phi$ values. $\mathcal{R}=0.003\%$ have been considered for the computation.}
\label{tab:det}
\end{table} 

A direct comparison between Fig.~\ref{fig:CS_BR} and Table~\ref{tab:det} shows the effect of ISR and a Gaussian BES on the detection prospects of a gauge-singlet scalar.
\subsection{Need for Radiative Return}
For muon colliders the integrated luminosity scales as $\mathbf{L}\sim\left(\frac{\sqrt{s}}{10~{\rm TeV}}\right)^2\times 10~{\rm ab}^{-1}$. Therefore, for example, if one considers a 3 TeV muon collider, the detection cross-section must be larger than 1 ab, i.e., $10^{-3}$ fb for the particle to be discovered. The results clearly suggest that the singlet scalar produced at the high-energy muon colliders following the proposed mechanism can be detected only if it hits the resonance. Moreover, the present numerical analysis corresponds to $\mathcal{O}(1)$ NP couplings. Therefore, the detection cross-section would be further suppressed if the leptonic constraints on $\xi_{L,R}^{i \mu}$ are invoked. For example, to satisfy the observed discrepancy in $(g-2)_\mu$, one has to consider $|\xi^{c \mu}|^2\sim\mathcal{O}(10^{-2})$ and $|\xi^{t \mu}|^2\sim\mathcal{O}(10^{-4})$~\cite{Mandal:2019gff}, resulting in an $\mathcal{O}(10^{-4})$ suppression of $\sigma_{XX}$ over the entire parameter space. However, the situation can be remarkably improved if one considers the radiative return~\cite{Chakrabarty:2014pja} where through the emission of a mono-chromatic photon from the initial states~($\mu$) $\phi$ can be resonantly produced at the muon colliders even if $\sqrt{s}>M_\phi$. In case of radiative return, a hard photon with energy $E_\gamma=\frac{(s-M_\phi^2)}{2\sqrt{s}}$ is emitted from the colliding muons. Thus, the effective COM energy is reduced, making it possible to explore a wider parameter space with a reasonable luminosity. For example, Ref.~\cite{Franceschini:2021aqd} has studied the production of 125 GeV SM Higgs at the TeV-scale muon colliders, where the Higgs is brought back to resonance through the radiative return. Though the analysis is not included here, one must consider radiative return to enhance the detectability of $\phi$ within the proposed analytical set-up. Such a study can be crucial in testing the present model as a viable mechanism to produce $\phi$ at the future multi-TeV muon colliders.

\section{Conclusion}
\label{sec:4}
\noindent
The present work studies a TeV-scale extension of the SM, where the NP interactions involve a scalar leptoquark $S_1$~($\bar{\mathbf{3}}$, {\bf 1}, 1/3) and a gauge-singlet scalar $\phi$. 
Though explaining the origin of $S_1$ in a GUT formulation is easy, the proposed model simply considers its interactions at the TeV scale. Within this minimal BSM framework, $\phi$ can effectively couple to the charged leptons, up-type quarks, gluons, and photons at one-loop level. Thus, through the effective $\bar{\mu}\mu\phi$ and $\gamma\gamma\phi$ couplings the SM-singlet state can be directly produced at the muon colliders without mixing with any Higgs-like doublet state. A detailed computation of the effective coupling strength $Y_{\bar{\mu}\mu\phi}$ has been presented as a function of the NP scale, followed by the discussion on various leading order decay modes of $\phi$. A sub-TeV SM-singlet scalar is observed to decay dominantly through the di-lepton channels, while the di-gluon mode becomes significant for the heavier scalars. In the proposed model, the production of $\phi$ follows from the muon fusion and photon fusion processes. Though muon fusion plays
the leading role at the resonance, for $|s-M^2_\phi|\neq 0$, it is negligibly suppressed. Thus, the photon fusion becomes significant to produce $\phi$ when $M_\phi$ is away from the resonance~(i.e., $\sqrt{s}>M_\phi$). However, the numerical results show that a notable production cross-section can be obtained only if $\sqrt{s}\to M_\phi$. Moreover, $\sigma_\phi$ would be further suppressed~(approximately by an $\mathcal{O}(10^{-4})$) if the low-energy leptonic constraints are imposed on the couplings. Therefore, one has to consider the radiative return to produce a sub-TeV SM-singlet scalar significantly at the high-energy muon colliders. The hard photon so emitted can be tagged with the existing decay modes of $\phi$, resulting in an interesting signature of a singlet scalar in the sub-TeV range. The decay modes $\phi\to \bar{u}u$~[$u\equiv u,\, c,\, t$] and $\phi\to \gamma\gamma$ being extremely suppressed in the considered parameter space, the detection probability of $\phi$ has been discussed only in the context of di-lepton and di-gluon channels. The results shown in Fig.~\ref{fig:CS_BR} correspond to the hard collision, whereas the results listed in Table~\ref{tab:det} encapsulate the effect of ISR and BES at the lowest order. Though the present work considers only the leading order contributions computed with $\mathcal{O}(1)$ NP couplings, the parameter space can be tuned through various phenomenological constraints to evade the currently available LHC bounds and significantly explain the detection prospects of $\phi$. The proposed analytical set-up, if augmented with the concept of radiative return, might be instrumental in predicting the discovery of the SM-singlet scalar at the future muon colliders.
 
\bigskip
\small \bibliography{Singlet_prod}{}
\bibliographystyle{JHEPCust}    
    
\end{document}